# Raman-probing the local ultrastrong coupling of vibrational plasmon-polaritons on metallic gratings

Rakesh Arul[1], Kishan Menghrajani[2], Marie S. Rider[2], Rohit Chikkaraddy[1], William L Barnes[2], and Jeremy J. Baumberg[1]

[1] NanoPhotonics Centre, Cavendish Laboratory, Department of Physics, JJ Thompson Avenue, University of Cambridge, Cambridge, CB3 0HE, United Kingdom
[2] Department of Physics and Astronomy, University of Exeter, Exeter, United Kingdom

ABSTRACT
**Strong coupling of molecular vibrations with light creates polariton states, enabling control over many optical and chemical properties. However, the near-field signatures of strong coupling are difficult to map as most cavities are closed systems. Surface-enhanced Raman microscopy of open metallic gratings under vibrational strong coupling enables the observation of spatial polariton localization in the grating near-field, without the need for scanning probe microscopies. The lower polariton is localized at the grating slots, displays a strongly asymmetric lineshape, and gives greater plasmon-vibration coupling strength than measured in the far-field. Within these slots, the local field strength pushes the system into the ultrastrong coupling regime. Models of strong coupling which explicitly include the spatial distribution of emitters can account for these effects. Such gratings form a new system for exploring the rich physics of polaritons and the interplay between their near- and far-field properties through polariton-enhanced Raman scattering (PERS).**

The interaction of light with ensembles of resonant two-level systems within a cavity of sufficient finesse can reach the strong coupling regime[1]. This manifests as a Rabi splitting of the original resonance into upper and lower polariton modes, besides uncoupled 'dark' states. Under strong coupling, many properties of the coupled emitter-photon system change, including chemical reactivities[2] and single photon nonlinearities[3]. Strong coupling has previously been achieved within single antenna-single emitter structures[4-6], high-finesse optical[3, 7] and microwave resonators[8], and more recently in the mid-infrared (IR) regime with vibrational strong coupling enabled by plasmonic metasurfaces[9-12]. The IR regime holds several advantages for strong coupling including large dipole oscillator strengths, easier cavity fabrication, but especially for direct access to molecular vibrations relevant to material properties. Typically, the energy and momentum of polaritons is probed through far-field scattering and emission spectroscopies. However, the requirement for high-finesse cavities implies that little light escapes outside the cavity resonances, limiting optical interrogation of near-field variations in polariton modes, which is expected to show rich physics. Probing the near-field of systems under strong coupling, which previously required challenging scanning near-field optical microscopies[13], constitutes an open problem in understanding light-matter interactions on sub-wavelength scales[14].

Surface plasmon polaritons (SPPs) at a metal-dielectric interface lie outside the light-line and cannot be excited by incident far-field radiation due to the momentum-mismatch. By structuring a metallic surface with a periodic grating, the optical wavevector acquires multiples of $k_g = 2\pi n_m/\Lambda$, for grating period $\Lambda$ and refractive index $n_m$ of the medium around the grating. At a metal-dielectric interface SPPs can thus be multiply scattered and form Bloch waves. Far-field radiation may then couple to the plasmonic grating-modes, and in the near-field can couple to molecular transitions ($\omega_0$) such as vibrational IR absorptions.

When an IR-active vibration is simultaneously Raman-active, polariton modes are also expected to appear in the Raman spectra in the ultrastrong coupling regime[15-17] (Fig. 1c). Despite such predictions, dark modes dominate the Raman scattering, preventing conclusive demonstrations of vibro-polaritonic Raman in microcavities[18-20] or open metasurfaces[12, 21], although phonon-polaritons can be seen in bulk crystal Restrahlen bands[22, 23]. Here, we show that surface-enhanced Raman (SERS) microscopy of micron-scale open gratings experiencing infrared vibrational strong coupling, which we term polariton-enhanced Raman spectroscopy (PERS), enables us to detect the localization and energy of polaritons in the near-field of the grating. This also eliminates the need for perturbing tip-based scanning probe microscopies.

PERS spectra of gratings under vibrational strong coupling are measured using a confocal Raman microscope (Fig. 1a) with a tightly focused (100x objective, 0.9 NA) 532 nm excitation laser (see SI Methods). The IR transmission spectra of the gratings display a clear anti-crossing of the polariton modes at normal incidence, when the $\Lambda$ = 4.7 µm grating mode strongly couples with the C=O stretch of PMMA ($\omega_0$ = 1732 cm$^{-1}$) at normal incidence (Fig. 1b). This matches a rigorous coupled wave analysis (RCWA) model (Fig. 1b). Fitting the Hopfield model yields a Rabi splitting $\Omega_{\text{Rabi}}$= 95 cm$^{-1}$, which exceeds the linewidths of the original plasmonic grating mode (45 cm$^{-1}$) and absorption at $\omega_0$ (SI Section S2).

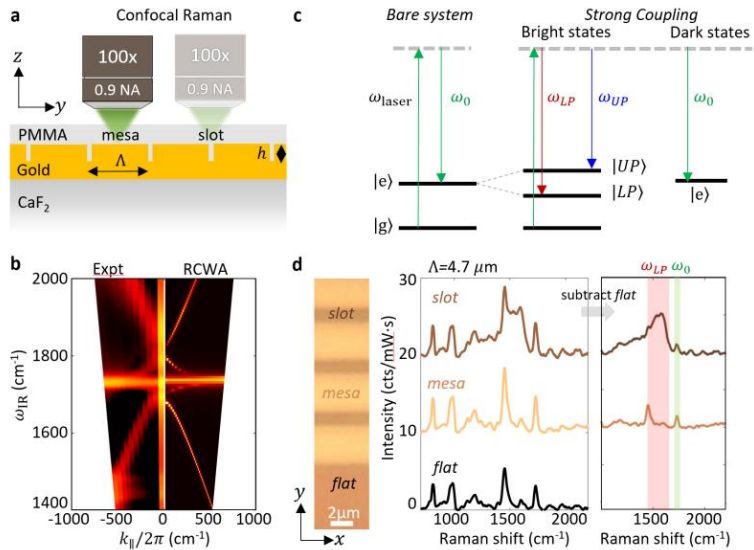

**Figure 1. Strong coupling in open gratings**. (**a**) Raman microscopy (pump 532 nm) of grating (period $\Lambda$=4.7 µm, slot width 1 µm, height $h$=0.1 µm, with 1 µm thick PMMA layer). (**b**) Grating IR reflectance spectra in experiment (expt, left) and theory (RCWA, right). (**c**) Raman scattering from ground state $|g\rangle$ to excited state $|e\rangle$ or PERS to lower $|LP\rangle$ and upper $|UP\rangle$ polariton states. (**d**) PERS spectra of slot and mesa

in grating compared to flat regions. Left: Optical image of grating. Middle: Background-corrected Raman spectra. Right: Raman spectra of slot and mesa after subtracting flat spectrum.

Open gratings allow Raman spectral mapping as a function of lateral position and height. This reveals three distinct Raman signatures: from the flat (reference) region, from the middle of the grating mesa, and from the grating slot (Fig. 1d). The flat region shows expected PMMA peaks, which are enhanced on the gold-coated mesa due to increased reflectivity from the gold. In the slots however, there is a clear extra contribution in the 1000-1600 cm$^{-1}$ region. Removing the background contribution reveals the spectral signature of this extra contribution (Fig. 1d right, brown), which is attributed to the lower polariton $|LP\rangle$ state (see Fig. 1c). The extracted $|LP\rangle$ spectrum is localized within the slot and displays an asymmetric shape peaked at ~1600 cm$^{-1}$ ($\omega_{LP}$). The possibility of molecular damage, as observed elsewhere[18], is discounted through intensity-dependent measurements that reveal different peaks associated with molecular damage while the original PMMA peaks remain unchanged (SI Section S4). No signal from the upper polariton mode $|UP\rangle$ is observed however. A broad electronic Raman scattering (ERS) or photoluminescence background is also present in the slot PERS spectra (SI Fig. S3c) due to excitation of edge plasmon modes.

**Spatial Raman mapping of LP**
The observed $|LP\rangle$ signature is enhanced and localized in the slot region of the $\Lambda$ = 4.7 µm grating, while the PMMA molecular band $\omega_0$ is only weakly enhanced (Fig. 2b,c,d, and SI Section S3). The $|LP\rangle$ signature is seen to be localized within the slots when focussing $z$=1 µm above the grating surface and resolved to be at the slot edges when focussing exactly on the surface ($z$=0 µm). This is more clearly visible in the laterally-averaged plots (Fig. 2e,g). The extracted energy separation between the peak of the $|LP\rangle$ and $\omega_0$ at each location in the slots reveals an average near-field coupling strength $g$ of 174 ± 30 cm$^{-1}$, much larger than measured in the far field through IR transmission (47.5 cm$^{-1}$).

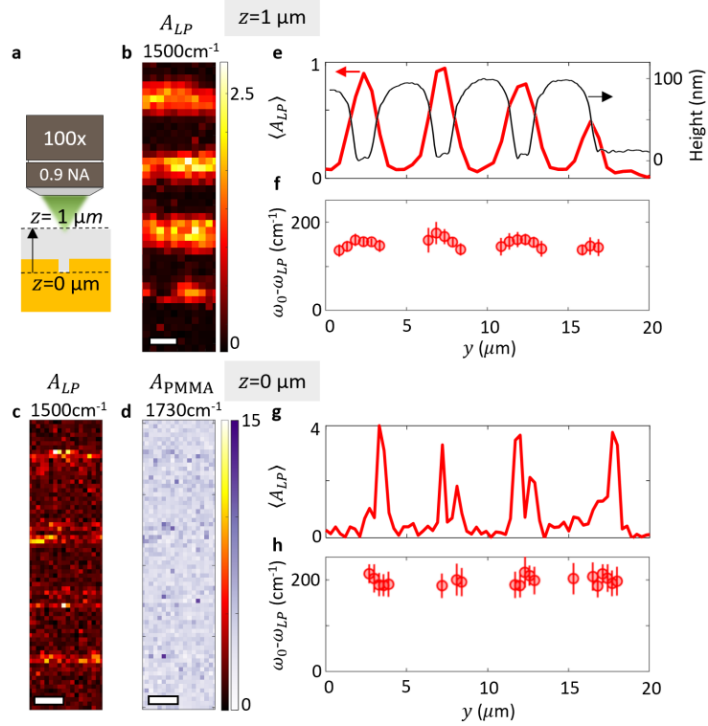

**Figure 2. Lower polariton enhanced Raman and localization in grating slots.** (**a**) Raman maps acquired at two different heights above the grating surface. (**b**) Raman map at PMMA surface ($z$=1 μm) showing integrated peak area of lower polariton ($A_{LP}$). (**c,d**) Raman maps at grating surface ($z$=0 μm) for (**c**) LP ($A_{LP}$) and (**d**) PMMA ($A_{PMMA}$) modes. White scale bars are 2μm. (**e-h**) Laterally averaged $A_{LP}$ and $\omega_{LP}$-$\omega_0$ across the grating for (**e,f**) $z$=1 μm and (**g,h**) $z$=0 μm.

To further confirm this assignment of $|LP\rangle$ polariton-enhanced Raman scattering (PERS), the dependence on grating mode detuning from $\omega_0$ is examined (Fig. 3). Changing the periodicity of the grating from 4 to 14 μm shifts this detuning (SI Section S5), as the position of the anti-crossing moves to higher wavevector, away from normal incidence (SI Fig. S10). This manifests as a reduced spectral weight and asymmetry in the $|LP\rangle$ PERS spectrum for Λ = 6 μm (Fig. 3a). For Λ = 13.8 μm, it is the second-order grating mode which now couples to $\omega_0$, however no clear PERS signal is observed, perhaps due to the greater sensitivity of the higher-order mode to fabrication imperfections. Only for anti-crossings at near normal-incidence is the strong $|LP\rangle$ PERS seen. Averaging the spectral area of the $|LP\rangle$ mode while scanning across the grating (Fig. 3b) confirms that it is always localized at the slots. The dependence of the intensity of $|LP\rangle$ PERS signals can be attributed to the different local field enhancements of the Raman excitation laser within the slots depending on the grating period (SI Section S6).

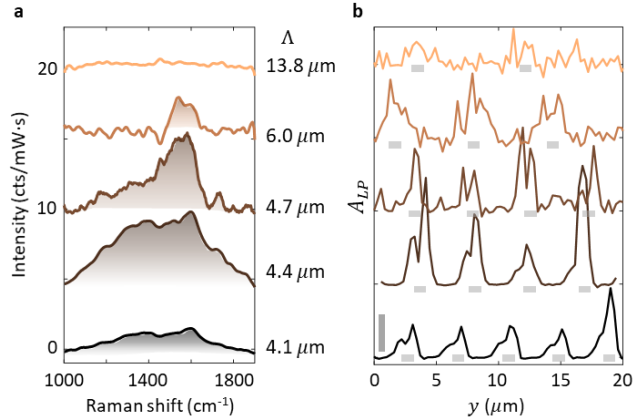

**Figure 3. Detuning of grating modes**. (**a**) PERS spectra of lower polariton mode for different grating periods (Λ). (**b**) Laterally-averaged lower polariton mode (integrated area) *vs y*-position, which is localised at the grating slots (vertical bar is 100 cts·cm$^{-1}$mW$^{-1}$s$^{-1}$). Horizontal gray bars indicate slot positions.

**Understanding localized polariton Raman scattering**

To explain the localization of the $|LP\rangle$ signal within the slots, the $|LP\rangle$ asymmetric PERS spectrum, the lack of an upper polariton signal, and the larger near-field Rabi splitting, we perform near-field electromagnetic simulations and develop a simplified theory for Raman scattering of polaritons within gratings. The $|LP\rangle$ PERS intensity depends on the localization of the infrared polariton (~1681cm$^{-1}$) and the visible Raman pump (532 nm). Finite-difference time-domain (FDTD) simulations show the $|LP\rangle$ optical near-field ($E_x$) is indeed tightly localized at the slot edges on the grating surface at $z$=0 μm (Fig. 4a). The 532 nm laser field is uniformly enhanced on the facet and at the slot edges (SI Fig. S14c). Hence, their combined impact is to out-couple the $|LP\rangle$ PERS signals more effectively from the slot edges.

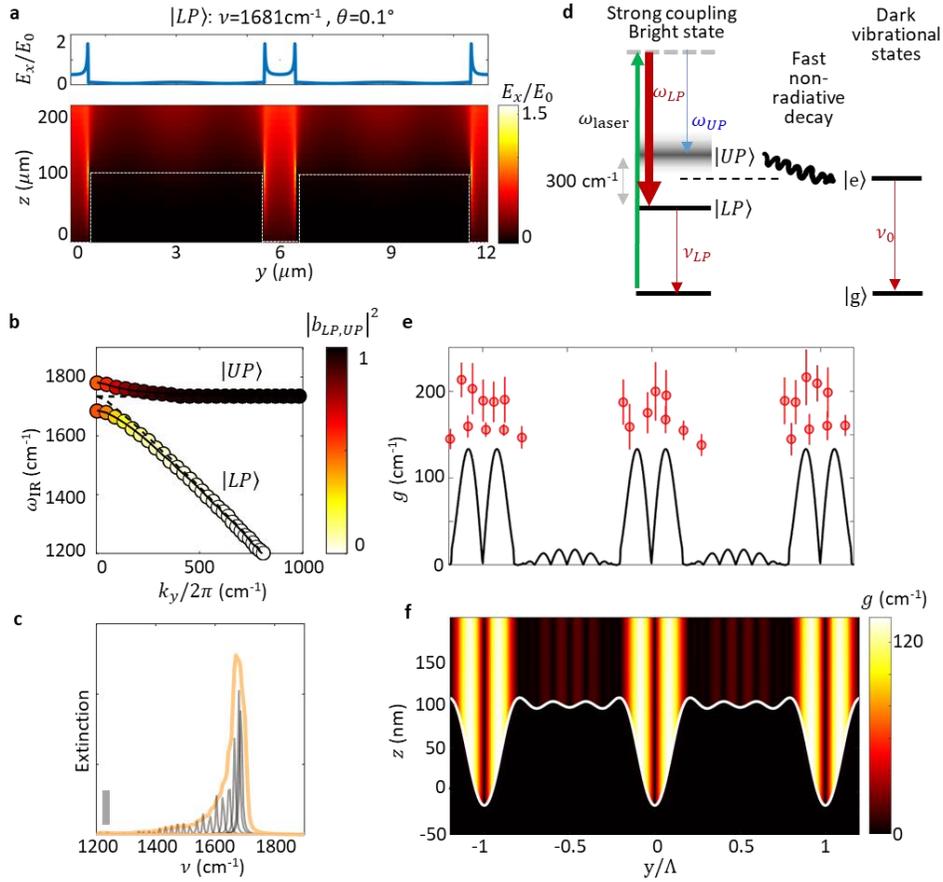

**Figure 4. Polariton-enhanced Raman scattering in gratings.** (a) Field distribution ($E_x/E_0$) for the lower polariton under near-normal incidence ($\theta$=0.1°) transverse-magnetic (TM) excitation perpendicular to the grating grooves (dashed). (b) Coupled oscillator fit to RCWA simulations of grating scattering for upper and lower polariton modes *vs* momentum ($k_y$). Colours show vibrational Hopfield coefficient fraction (SI Section S2). (c) Molecular density-of-states (corrected and angle-averaged) showing asymmetric broadening of LP peak (grey bar indicating extinction of 0.1). (d) Polariton states showing scattering from ground state to bright (strongly coupled) and dark states, labelling Raman scattering ($\omega$) and infrared absorption ($\nu$). (e) Plasmon-vibration coupling strength $g$ *vs* normalised position $y/\Lambda$, with red points from Fig. 2. (f) Map of spatial plasmon-vibration coupling strength $g$ along the $\Lambda$=4.7 μm grating (modelled surface profile in white).

The asymmetry of the $|LP\rangle$ PERS signal is likely due to angle averaging inherent in confocal Raman microscopy with a high numerical-aperture objective to get spatial profiles. The visible-frequency Raman dipoles sample the $|LP\rangle$ across a range of momentum $k_y$ in the infrared (Fig. 1b), which broadens the range of frequencies measured. The fraction of molecular oscillator in each hybrid polariton state changes with momentum $k_y$, and is quantified by the Hopfield coefficient $b_{LP,UP}$ as a function of angle for the $|LP\rangle$ and $|UP\rangle$ (Fig. 4b). The molecular fraction contributes to PERS, so using density-of-states estimates[18], we weight each simulated reflectance spectrum (SI Section S3) with its molecular fraction and evaluate the sum up to the numerical aperture of our objective lens. The resulting spectrum is asymmetric (Fig. 4c) with a significant low-frequency tail as observed in experiments (Fig. 1d). We note this approximation only considers the $k_y$ momentum component and not the full parabolic dispersion including $k_x$. By contrast, the $|UP\rangle$ PERS signal is suppressed due to fast non-radiative decay of the upper polaritons into the reservoir of dark vibrational states present in the system, and the lower molecular fraction of $|UP\rangle$ states at small $k_y$ (Fig. 4b). This is analogous to electronic strong coupling, where only emission from the lower polariton branch is detected at room temperature[24].

Finally, the experimentally observed enhanced near-field Rabi splitting (distinct from far-field) is recovered in our theoretical model showing spatial variation in molecule-plasmon coupling constant $g$ due to spatial variation in the optical field strength. The dominant contribution to the lower and upper polariton optical fields comes from the coupled first-order ($\pm 2\pi/\Lambda$) grating-scattered branches of the surface plasmon[25]. Using the Chandezon method to calculate the contribution to the optical field from the lower grating-scattered branch at $k_y = 0$ from a rectangular grating (SI Section S7), we find the spatially dependent light-molecule coupling from, $g_i(y,z) \approx -\mu_i \cdot \boldsymbol{E}(y,z)$, where $g_i$ is the plasmon-vibration coupling strength for molecule $i$ in position $y$ on the grating at height $z$ from the gold surface, $\boldsymbol{E}$ is the optical field associated with the lower grating-scattered branch at $k_y = 0$, and $\mu_i$ is the vibration IR transition dipole moment. While the effective coupling strength[26] $\langle g \rangle = \sqrt{\sum_i g_i^2}$ is measured in far-field extinction spectra, in the near-field molecules experience a spatially-varying $g_{\mathrm{NF}}$ which is maximum near slot edges and greatly reduced on the mesa (Fig. 4e-f), directly seen in the PERS spectra of the $|LP\rangle$ state, and can greatly exceed the far-field mean $\langle g \rangle$. The value of $g_{\mathrm{NR}}$ has an upper bound given by the bulk optical parameters of PMMA[27]. We find that this maximum value ~ 140 cm$^{-1}$ (SI Section S8) is consistent with the maximum near-field value we find here of 174 ± 30 cm$^{-1}$. Thus, any near-field enhancement of the strong coupling effect can at most place the LP at ~1590 cm$^{-1}$, broadly consistent with our observations. We note that this bulk limit may help resolve the thus-far puzzling observations of Shalabney et al.[20] and Menghrajani et al.[9], both of whom found Raman features at ~1590 cm$^{-1}$; both might be due to Raman signals associated with localised modes. This highlights the role of (i) the visible wavelength SERS effect in amplifying signals from a small number of molecules at edges, and (ii) localized hot-spots in the infrared that couple to the dispersive grating mode giving rise to the large coupling strength $g$ measured in PERS. The large near-field $g_{\mathrm{NF}}$ = 174 ± 30 cm$^{-1}$ pushes the system into the ultrastrong coupling regime ($\omega_{\mathrm{Rabi}}/\omega_0$=0.2>0.1), with clear effects on Raman spectra as predicted[16, 17].

Open grating systems thus allow the direct observation of Raman signatures from lower polaritons of systems under vibrational strong coupling. Grating systems uniquely allow a visible laser probe to map out the local density-of-polaritonic-states via Raman scattering. These signatures are not observed for systems out of strong coupling, nor for different molecular vibrations which are uncoupled to the grating cavity modes. As the collective Rabi splitting ($\Omega$) depends on the number of molecules ($N$) as $\Omega \propto N^{1/2}$, the individual splitting experienced by a molecule depends on $N^{-1/2}$. Hence, a small mode volume in the IR (encompassing small $N$) is critical to observe splitting effects in polariton-enhanced Raman scattering. For previous measurements in microcavities and plasmonic surface lattice resonators[9-11], $N$ was much larger since mode volumes were ~$\lambda^3$. Within gratings, the number of molecules involved is much reduced due to the tight localization of the field at the grating slot edges, whose signals are effectively enhanced by SERS. The degree of local field confinement is such that the molecules at the edges are in the ultrastrong coupling regime ($\omega_{\text{Rabi}}/\omega_0 = 0.2$).

ASSOCIATED CONTENT

**Supporting Information**.

AUTHOR INFORMATION

**Corresponding Author**

* Prof Jeremy J Baumberg, jjb12@cam.ac.uk

* Prof William L Barnes, W.L.Barnes@exeter.ac.uk

ACKNOWLEDGMENT

We acknowledge support from European Research Council (ERC) under Horizon 2020 research and innovation programme THOR (Grant Agreement No. 829067) and PICOFORCE (Grant Agreement No. 883703). R.A. acknowledges support from the Rutherford Foundation of the Royal Society Te Apārangi of New Zealand, and the Winton Programme for the Physics of Sustainability. R.C. and R.A. acknowledge support from Trinity College, University of Cambridge. K.S.M. acknowledges financial support from the Leverhulme Trust research grant 'Synthetic biological control of quantum optics'.